\documentstyle[sprocl]{article}

\include{psfig}

\def\Journal#1#2#3#4{{#1} {\bf #2}, #3 (#4)}

\def\NPA{{\em Nucl. Phys.} A}
\def\JPG{{\em J. Phys.} G}
\def\PLB{{\em Phys. Lett.} B}

\def\PRC{{\em Phys. Rev.} C}

\begin{document}

\title{$\pi NN^*(1440)$ AND $\sigma NN^*(1440)$
COUPLING CONSTANTS FROM A MICROSCOPIC $NN \to NN^*(1440)$ POTENTIAL}

\author{P. GONZ\'ALEZ}

\address{Dpto. de F\' \i sica Te\'orica and IFIC\\
Universidad de Valencia - CSIC, E-46100 Burjassot, Valencia, Spain\\
E-mail: pedro.gonzalez@uv.es}

\author{B. JULI{\'A}-D{\'I}AZ, A. VALCARCE, F. FERN\'ANDEZ}

\address{Grupo de F\' \i sica Nuclear \\
Universidad de Salamanca, E-37008 Salamanca, Spain\\E-mail: 
bjulia@usal.es, valcarce@mozart.usal.es, fdz@usal.es}

\maketitle
%\abstracts{By means of a $NN \to NN^*(1440)$ transition
%potential derived in a parameter-free way from a $NN$ quark based potential,
%we determined simultaneously the $\pi NN^*(1440)$ and $\sigma NN^*(1440)$
%coupling constants.} 

The $N^*(1440)$, or Roper resonance, plays an important role in
nucleon and nuclear dynamics as an intermediate state. The excitation of a
Roper may mediate pion electro- and photoproduction processes~\cite{uno}, pion and
double pion production in $NN$ reactions~\cite{dos,tre}, heavy ion 
collisions~\cite{cua}, etc. In this context, the knowledge
of the $NN \to NN^*(1440)$ interaction and the 
$\pi NN^*(1440)$ and $\sigma NN^*(1440)$ coupling constants should be of great help.

The usual way to determine meson$-NN$ coupling constants is trough the fitting
of $NN$ scattering data with phenomenological meson exchange
models. Therefore, a consistent way to obtain
meson$-NN^*(1440)$ coupling constants is from a transition 
$NN \to NN^*(1440)$ potential, in particular
when ratios over meson$-NN$ coupling constants are to be considered. In order to
derive the transition potential we shall follow the same quark model
approach previously used for $NN$ scattering~\cite{cin,sei}, this is
we shall start from a quark-quark ($qq$)
interaction containing confinement, one-gluon
exchange (OGE), one-pion exchange (OPE) and one-sigma exchange (OSE)
terms and carry out a Born-Oppenheimer approximation. The quark
treatment presents two main advantages, on the one hand once the parameters
of the $qq$ potential are fixed from $NN$ data, there is not
any free parameter, on the other hand it allows all the baryon-baryon
interactions to be dynamically considered on an equal footing (actually the
same framework has been applied to the $\Delta$ excitation case~\cite{sie}).

Explicitly, the $NN\rightarrow NN^*(1440)$ potential at interbaryon
distance $R$ is obtained by sandwiching the $qq$ potential, $V_{qq}$,
between $NN$ and $NN^*(1440)$ states, written in terms of quarks, for all
the pairs formed by two quarks belonging to different baryons.
The $qq$ potential has been very much detailed elsewhere~\cite{cin,sei}. It
reads:
\begin{equation}
V_{qq}(\vec{r}_{ij})=V_{CON}(\vec{r}_{ij})+V_{OGE}(\vec{r}_{ij})+V_{OPE}(
\vec{r}_{ij})+V_{OSE}(\vec{r}_{ij})\,
\end{equation}
where $\vec{r}_{ij}$ is the interquark distance. $V_{CON}$ is the confining
potential taken to be linear ($r_{ij}$) and $V_{OGE}$ is the usual
perturbative one-gluon-exchange interaction containing Coulomb ($\frac{
1}{r_{ij}}$), spin-spin (${\vec{\sigma}}_{i}\cdot {\vec{\sigma}}_{j})$ and
tensor ($S_{ij}$) terms. For future purposes we detail the central part
of the one-pion, $V_{OPE}$, and one-sigma, $V_{OSE}$, exchange interactions given by:

\begin{equation}
V_{OPE} ({\vec r}_{ij}) = {\alpha_{ch} \over 3} {\frac{\Lambda^2 
}{\Lambda^2 - m_\pi^2}} \, m_\pi \left[ Y (m_\pi \, r_{ij}) - 
{\frac{ \Lambda^3 }{m_{\pi}^3}} \, Y (\Lambda \, r_{ij}) \right] {\vec \sigma
}_i \cdot {\vec \sigma}_j \, {\vec \tau}_i \cdot {\vec \tau}_j \, ,
\end{equation}

\begin{equation}
V_{OSE}({\vec{r}}_{ij})=-\alpha _{ch}\,{\frac{4\,m_{q}^{2}}{m_{\pi }^{2}}}{%
\frac{\Lambda ^{2}}{\Lambda ^{2}-m_{\sigma }^{2}}}\,m_{\sigma }\,\left[
Y(m_{\sigma }\,r_{ij})-{\frac{\Lambda }{{m_{\sigma }}}}\,Y(\Lambda \,r_{ij})%
\right] \,,
\end{equation}
where $\alpha _{ch}$ is the chiral coupling constant and $\Lambda $
is a cutoff parameter.
The values chosen for the parameters are taken from Ref.~\cite{sei}.

The $N^*(1440)$ and $N$ states are given in terms of quarks by
$|N^*(1440)\rangle =\left\{ \sqrt{\frac{2}{3}}|[3](0s)^{2}(1s)\rangle -
\sqrt{\frac{1}{3}}|[3](0s)(0p)^{2}\rangle \right\} \otimes \lbrack 1^{3}]_{c}$
and $|N\rangle =|[3](0s)^{3}\rangle \otimes \lbrack 1^{3}]_{c}$ where
$[1^{3}]_{c}$ is the completely antisymmetric color state, $[3]$ is the
completely symmetric spin-isospin state and $0s$, $1s$, and $0p$, stand for
harmonic oscillator orbitals.

The transition potential obtained can be written at all
distances in terms of baryonic degrees of freedom~\cite{och}. One should
realize that a $qq$ spin and isospin independent potential as for instance
the scalar OSE, gives rise at the baryon level, apart from a
spin-isospin independent potential, to a spin-spin, an isospin-isospin and a
spin-isospin dependent interactions~\cite{cin}. Nonetheless for
distances $R\geq 4$ fm, where quark antisymmetrization interbaryon effects
vanish, we are only left with the direct part, i.e. with a scalar OSE
at the baryon level. The same kind of arguments can be applied to
the OPE potential. Thus asymptotically ($R\geq 4$ fm) OSE and
OPE have at the baryon level the same spin-isospin structure than
at the quark level. 
Hence we can parametrize the asymptotic central interactions as (the
$\Lambda $ depending exponential term is negligible asymptotically as
compared to the Yukawa term) 
\begin{eqnarray}
V_{NN\rightarrow NN^*(1440)}^{OPE}(R) & = & \frac{1}{3}\,\frac{g_{\pi NN}}
{\sqrt{4\pi }}\,\frac{g_{\pi NN^{\ast }(1440)}}{\sqrt{4\pi }}\,\frac{m_{\pi}
}{2M_{N}}\,\frac{m_{\pi }}{2(2M_{r})}\,\frac{\Lambda ^{2}}{\Lambda
^{2}-m_{\pi }^{2}} \nonumber \\
 & & [(\vec{\sigma}_{N}.\vec{\sigma}_{N})(\vec{\tau}_{N}.\vec
{\tau}_{N})]\,\frac{e^{-m_{\pi }R}}{R}\,,
\label{eq:lrg}
\end{eqnarray}
and
\begin{equation}
V_{NN\rightarrow NN^*(1440)}^{OSE} (R)=- \, \frac{g_{\sigma NN}}{\sqrt
{4\pi }} \, \frac{g_{\sigma NN^{\ast}(1440)}}{\sqrt{4\pi }} \, \frac{\Lambda
^{2}}{\Lambda ^{2}-m_{\sigma }^{2}} \, \frac{e^{-m_{\sigma }R}}{R} \, ,
\label{eq:slrg}
\end{equation}
where $g_{i}$ stands for the coupling constants at the baryon
level and $M_{r}$ is the reduced mass of the $NN^*(1440)$ system.
By comparing these baryonic potentials with the asymptotic behavior of the
OPE and OSE previously obtained from the quark calculation we can extract
the $\pi NN^*(1440)$ and $\sigma NN^*(1440)$ coupling constants.

The $\Lambda^2/(\Lambda^2-m_i^2)$ vertex factor
in expressions (\ref{eq:lrg}) and (\ref{eq:slrg}) 
comes from the vertex form factor chosen in
momentum space as a square root of monopole $\left( \frac{\Lambda ^{2}}{
\Lambda ^{2}+\vec{q}^{\,\,2}}\right) ^{\frac{1}{2}}$, the same choice taken
at the quark level, where chiral symmetry requires the same form for pion
and sigma. A different choice for the form factor at the baryon level,
regarding its functional form as well as the value of $\Lambda$, would give
rise to a different vertex factor and eventually to a different functional
form for the asymptotic behavior.
Then, the extraction from any model of the
meson-baryon-baryon coupling constants depends on this choice. We shall say
they depend on the coupling scheme.

For the OPE and for our value of $\Lambda =4.2$ fm$^{-1}$,
$\frac{\Lambda ^{2}}{\Lambda ^{2}-m_{\pi }^{2}}=1.0286$, pretty close to 1.
As a consequence, in this case the use of our form factor or a modified
monopole form at baryonic level makes little difference in the determination
of the coupling constant. This fact is used when fixing $\frac{g_{\pi qq}^{2}
}{4\pi }$ from the experimental value of $\frac{g_{\pi NN}^{2}}{4\pi }$
extracted from $NN$ data. The value we use for $\alpha _{ch}=\frac{m_{\pi
}^{2}}{4m_{q}^{2}}\frac{g_{\pi qq}^{2}}{4\pi }=\left( \frac{3}{5}\right) ^{2}
\frac{g_{\pi NN}^{2}}{4\pi }\frac{m_{\pi }^{2}}{4m_{N}^{2}}e^{-\frac{m_{\pi
}^{2}b^{2}}{2}}=0.027$ corresponds to $\frac{g_{\pi NN}^{2}}{4\pi }=14.8$.

To get $g_{\pi NN^*(1440)}$ we turn to our
numerical results for the $^1S_0$ OPE potential and fit its asymptotic
behavior (in the range $R:5\rightarrow 9$ fm) to Eq. (\ref{eq:lrg}). We obtain
\begin{equation}
\frac{g_{\pi NN}}{\sqrt{4\pi }} \frac{g_{\pi NN^{\ast}(1440)}}{\sqrt{4\pi }} 
\frac{\Lambda ^{2}}{\Lambda ^{2}-m_{\pi }^{2}}= \, - \, 3.73 \, ,
\end{equation}
i.e. $\frac{g_{\pi NN^*(1440)}}{\sqrt{4\pi }}=-0.94$. Let us
note that the sign comes out from the arbitrary choice of the overall phase
of the $N^*(1440)$ wave function with respect to the $N$ wave function.
Hence it is more appropriate to quote the absolute value. Furthermore the
coupling scheme dependence can be explicitly eliminated if we compare
$g_{\pi NN^{\ast }(1440)}$ with $g_{\pi NN}$ extracted from the
$NN\rightarrow NN$ potential within the same quark model approximation. Thus
we get
\begin{equation}
\left| \frac{g_{\pi NN^{\ast }(1440)}}{g_{\pi NN}}\right| =0.25\,.
\end{equation}
The value obtained for this ratio is similar to that obtained in Ref.~\cite{nue}
and a factor 1.5 smaller than the one obtained from the analysis of
the partial decay width.
By proceeding in the same way for the OSE potential we have
\begin{equation}
\left| \frac{g_{\sigma NN^{\ast }(1440)}}{g_{\sigma NN}}\right| =0.475\,.
\end{equation}
This result agrees quite well with the only experimental available result,
obtained in Ref.~\cite{die} from the fit of the cross section of the
isoscalar Roper excitation in $p(\alpha ,\alpha ^{\prime })$ in the 10$-$15
GeV region.

Furthermore, we can give a very definitive prediction of the magnitude and
sign of the ratio of the two ratios,

\begin{equation}
\frac{g_{\pi NN^{\ast}(1440)}}{g_{\pi NN}}=0.53 \, \frac{g_{\sigma
NN^{\ast}(1440)}}{g_{\sigma NN}} \, ,
\end{equation}
which is an exportable prediction of our model.

We should finally notice that for dynamical applications our results should
be implemented including the $N^*(1440)$ width. Quantum
fluctuations of the two baryon center-of-mass, neglected within the
Born-Oppenheimer approach, could also play some role. Though these
improvements will have a quantitative effect we think, as suggested by the
values we get, our predictions will not be very much modified. In this sense
they could serve either as a first step for more refined calculations or as
a possible guide for phenomenological applications.

\section*{Acknowledgments}
We thank to E. Oset for suggesting discussions.
This work has been partially funded by Junta de
Castilla y Le\'{o}n under Contract No. SA-109/01, and by EC-RTN,
Network ESOP, Contract HPRN-CT-2000-00130.


\section*{References}
\begin{thebibliography}{99}
\bibitem{uno} H. Garcilazo and E. Moya de Guerra, \Journal{\NPA}{562}{521}{1993}.

\bibitem{dos} M.T. Pe\~na {\it et al}, \Journal{\PRC}{60}{045201}{1999}.

\bibitem{tre} L. Alvarez-Ruso, \Journal{\PLB}{452}{207}{1999}.

\bibitem{cua} B.A. Li {\it et al}, \Journal{\PRC}{50}{2675}{1994}.

\bibitem{cin} F. Fern\'{a}ndez {\it et al},
\Journal{\JPG}{19}{2013}{1993}.

\bibitem{sei} D.R. Entem {\it et al}, \Journal{\PRC}{62}{034002}{2000}.

\bibitem{sie} A. Valcarce {\it et al}, \Journal{\PRC}{49}{1799}{1994}.

\bibitem{och} K. Holinde, \Journal{\NPA}{415}{477}{1984}.

\bibitem{nue} D.O. Riska and G.E. Brown, \Journal{\NPA}{679}{577}{2001}.

\bibitem{die} S. Hirenzaki {\it et al},
\Journal{\PLB}{378}{29}{1996}; \Journal{\PRC}{53}{277}{1996}.

\end{thebibliography}
\end{document}